\documentclass[twocolumn,preprintnumbers,amsmath,amssymb,prl]{revtex4}

\usepackage{graphicx}
\usepackage{dcolumn}
\usepackage{bm}
\usepackage{color}

\begin{document}

\preprint{}

\title{Non-thermal photocoercivity effect in a ferromagnetic semiconductor}

\author{G.~V.~Astakhov$^{1}$}
\altaffiliation{Also at A.~F.~Ioffe Physico-Technical Institute,
RAS, 194021 St. Petersburg, Russia} \email[\\ E-mail:
]{astakhov@physik.uni-wuerzburg.de}
\author{H.~Hoffmann$^{1}$}
\author{V.~L.~Korenev$^{2}$}
\author{T.~Kiessling$^{1}$}
\author{J.~Schwittek$^{1}$}
\author{G.~M.~Schott$^{1}$}
\author{C.~Gould$^{1}$}
\author{W.~Ossau$^{1}$}
\author{K.~Brunner$^{1}$}
\author{L.~W.~Molenkamp$^{1}$ }

\affiliation{ $^{1}$Physikalisches Institut (EP3),
Universit\"{a}t W\"{u}rzburg, 97074 W\"{u}rzburg, Germany \\
$^{2}$A.F.Ioffe Physico-Technical Institute, Russian Academy of
Sciences, 194021 St.Petersburg, Russia}

\date{\today}

\begin{abstract}
We report a photoinduced change of the coercive field, i.e., a
photocoercivity effect (PCE), under very low intensity
illumination of a low-doped (Ga,Mn)As ferromagnetic semiconductor.
We find a strong correlation between the PCE and the sample
resistivity. Spatially resolved dynamics of the magnetization
reversal rule out any role of thermal heating in the origin of
this PCE, and we propose a mechanism  based on the light-induced
lowering of the domain wall pinning energy. The PCE is local and
reversible, allowing writing and erasing of magnetic images using
light.
\end{abstract}

\pacs{78.20.Ls, 75.60.-d, 75.50.Pp}

\maketitle

Magneto-optical (MO) recording techniques currently attract much
interest \cite{HAMR, app1, app3, app5, app2, app4} due to the
non-volatility, low cost, and the removability of media they
offer. Traditionally, the light is used to modify the strength of
magnetic interaction. Because a very large number of magnetic ions
is essential to achieve ferromagnetism the intensity of the light
needed to be rather high. This results in heating of the recording
media, regardless whether or not the thermomagnetic effect is the
exploited physical mechanism. The resulting heat dissipation is an
obviously undesirable side effect which leads to degradation of
the recording media \cite{heating} and wastes significant
resources. Here, we demonstrate a concept for MO recording which
circumvents this problem by focusing our action on the de-pinning
of domain walls, instead of trying to modify the magnetic
interaction strength. The pinning efficiency can be very sensitive
to the light even of low intensities, because the concentration of
domain wall pinning centers is much lower than that of magnetic
ions. Furthermore, in contrast to previous works on light assisted
magnetization reversal \cite{InMnAs}, such a photoinduced change
of the coercive field, i.e., a photocoercivity effect \cite{UFN}
(PCE), is local and reversible. This provides an approach to
non-thermal, low-power MO recording.

Our experiments are performed on a prototype system in the form of
a low-doped (Ga,Mn)As ferromagnetic semiconductor
\cite{FerroSemi}. The sample used in this study is a
$0.36$-$\mathrm{\mu m}$-thick $\mathrm{Ga_{1-x}Mn_{x}As}$ layer
(nominally $x \approx 0.01$), grown by low-temperature (LT)
molecular beam epitaxy (MBE) \cite{MBE} on a (001)-oriented GaAs
substrate and LT-GaAs buffer. The sample shows partial
compensation of Mn p-doping resulting in an insulating transport
character at low temperatures. SQUID measurements show an average
Mn concentration $x \approx 0.005$ and a Curie temperature $T_C =
25$~K. Because of the low hole density, the sample exhibits a
clear perpendicular-to-plane magnetic anisotropy (PMA) at low
temperatures \cite{Sawicki}.

\begin{figure}[tbp]
\includegraphics[width=.39\textwidth]{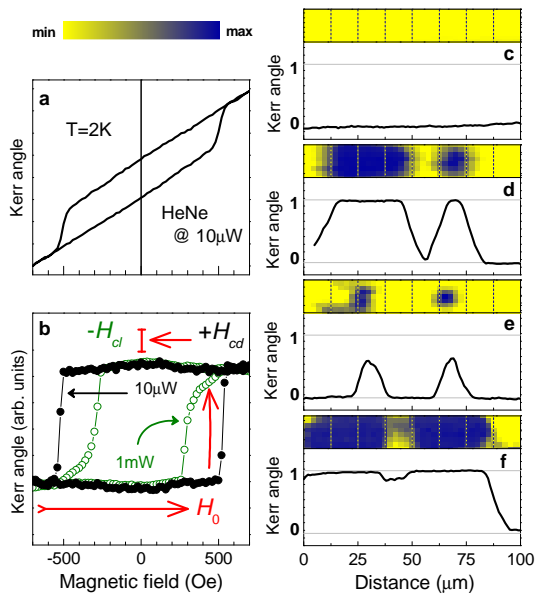}
\caption{ (a) MOKE hysteresis loops probed by a HeNe laser. (b)
MOKE hysteresis loops recorded for two powers of a NeHe laser
after subtracting the linear background. Arrows indicate the
light-assisted magnetization reversal process in an external
magnetic field $H_0$. (c) Zero-field magnetization profile
obtained after initialization in a negative field ($-1$~kOe). Byte
\{00000000\}. (d) as (c) but following a write procedure
(performed by 1 ms pulses from a HeNe laser with $P =
100$~$\mathrm{\mu W}$). Byte \{01110100\}. (e) as (d) following a
partial deletion procedure. Byte \{00100100\}. (f) as (e)
following a rewrite. Byte \{11101110\}. Thick lines in lower parts
of (c)-(f) are cross sections of the magnetization profiles. }
\label{fig1}
\end{figure}

Magnetic hysteresis loops are recorded by means of the
magneto-optical Kerr effect (MOKE). The angle of the Kerr rotation
$\theta$ is proportional to the perpendicular-to-plane component
of the magnetization $M$, $\theta \propto  M$. In our experiments
we use a HeNe laser ($\hbar \omega = 1.96$~eV) with intensity
varying from 1~$\mathrm{\mu W}$ to 1~mW. The linearly polarized
laser light, modulated at a frequency of 100~kHz by a
photo-elastic modulator, is focused to a spot size of about
$10$~$\mathrm{\mu m}$. The polarization of the reflected beam
($\theta$) is detected by a balanced photodiode scheme and a
lock-in amplifier. In the MO recording experiments of
Figs.~\ref{fig1}(c)-(f), a microscope objective with a numerical
aperture 0.42 (providing a spatial resolution of a few microns) is
utilized to focus the HeNe laser beam. The focusing lens is
mounted on a piezo system with submicron resolution, allowing to
record a surface scan of the magnetization profile. Where
additional illumination of the sample is needed, we use a Xenon
lamp, selecting the appropriate wavelength with a monochromator.
This radiation is defocused to yield a 1-mm diameter spot on the
sample surface. The experiments are performed in a flow cryostat
allowing for temperature dependent experiments, with the sample
immersed in superfluid helium for measurements done at $T = 2$~K.
An external magnetic field of up to 1~kOe can be applied using an
electromagnet mounted outside the cryostat.

Figure~\ref{fig1}(a) shows a typical MOKE hysteresis loop for the
(Ga,Mn)As sample, where the external magnetic field is applied
perpendicular to the sample plane. It is taken with an attenuated
HeNe laser beam at very low power ($P = 10$~$\mathrm{\mu W}$).
Figure~\ref{fig1}(b) (filled circles) shows the same data after
removing the linear background introduced by the cryostat windows.
We find that at these low light levels, the 'dark' coercivity $
H_{cd} = 525$~Oe. When the illumination power is increased to $P =
1$~mW (open circles), the coercive field of the region under
illumination is reduced by about 40\%, yielding a coercivity under
illumination $ H_{cl} = 285$~Oe. The difference between $H_{cd}$
and $H_{cl}$ is the PCE, and, as we will show below, it leads to
the formation of a local domain of magnetization opposite to that
of the bulk. The effect is reversible, i.e. when the power of the
HeNe laser is reduced back to 10~$\mathrm{\mu W}$, the 'dark'
hysteresis loop is recovered.

This controlled formation of localized magnetization domains
enables a complete MO recording cycle. To illustrate this, we
define an area of $15 \, \mathrm{\mu m} \times 100 \, \mathrm{\mu
m}$ on the sample surface to a byte, consisting of 8 side-by-side
bits. We define a bit with magnetization pointing down (up) to
correspond to a logical 0 (1). The (Ga,Mn)As layer is initially
prepared to be uniformly magnetized down by the application of a
sufficiently strong negative magnetic field [Fig.~\ref{fig1}(c)],
yielding a \{00000000\} byte . A magnetic field $H_0= + 470$~Oe,
which is weaker than the coercive field in the dark but stronger
than the coercive field in light, $H_{cl}< H_0 <H_{cd}$ is then
applied. Local illumination of the sample will now induce a
transition [vertical arrow in Fig.~\ref{fig1}(b)] into the state
with magnetization directed upwards. Scanning the laser beam over
the sample surface allows the writing of any desired image, which
is retained even when the external field is switched off [e.g., a
byte \{01110100\} is written in Fig.~\ref{fig1}(d)]. In a similar
manner, the data can be completely or partially erased [e.g., the
byte \{00100100\} in Fig.~\ref{fig1}(e)] by local illumination in
a magnetic field ($H_0= - 470$~Oe) applied in the opposite
direction. Missing bits can be added to a stored image at a later
time [byte \{11101110\} in Fig.~\ref{fig1}(f)].

\begin{figure}[tpb]
\includegraphics[width=.37\textwidth]{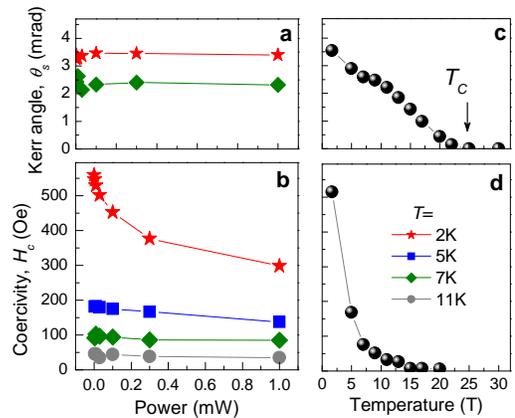}
\caption{Kerr angle in saturation $\theta_s$ as a function of (a)
illumination power (at $T = 2$~K  and $T = 7$~K) and (b)
temperature (for $P = 10$~$\mu W$). Coercive field $H_c$ (c) as a
function of illumination power at different temperatures ($T =
\,$2, 5, 7 and 11~K), and (d) as a function of temperature ($P =
10$~$\mu W$). In all panels, the solid lines are guides to the
eye.} \label{fig2}
\end{figure}

Next, we address the physics of the PCE. The MOKE hysteresis loops
are characterized by two key parameters: The Kerr angle in
saturation ($\theta_s$) and the coercive field $H_c$. These
parameters can be easily extracted, respectively, from the
half-height and half-width of the opening in the hysteresis loop.
Figure~\ref{fig2} shows the dependence of $\theta_s$ and $H_c$ on
illumination power ($P$) and temperature ($T$). The Kerr angle
decreases monotonically with rising temperature
[Fig.~\ref{fig2}(c)] and is nearly independent of the illumination
power [Fig.~\ref{fig2}(a)].

\begin{figure}[tpb]
\includegraphics[width=.39\textwidth]{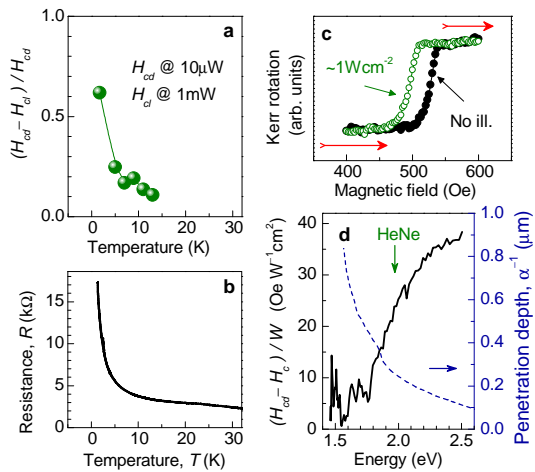}
\caption{(a) The light-induced change in coercive field
$(H_{cd}-H_{cl})$ normalized to the coercivity in dark $H_{cd}$.
(b) Resistance versus temperature. (c) Field-induced switching
measurement with (open symbols) and without (solid symbols)
additional illumination from a Xenon lamp with an intensity of
about $1$~$\mathrm{W \, cm^{-2}}$. (d) Difference in coercive
field scaled by the illumination intensity as function of photon
energy, shown along with the characteristic penetration depth
$\Lambda = \alpha ^{-1}$ of the light into GaAs. } \label{fig3}
\end{figure}

Our crucial observation is the reduction of the coercive field as
a function of the illumination power $H_c (P)$ as shown in
Fig.~\ref{fig2}(b). At low temperature ($T = 2$~K) the effect is
rather strong with the coercivity reducing by nearly 25\% before
showing a saturation tendency at $P > 300$~$\mathrm{\mu W}$. A
photoinduced change in the coercive field, $H_{cd} - H_c(P)$, at
these power levels is a rather unexpected and nontrivial behavior.
The PCE disappears rapidly with increasing temperature [see
Fig.~\ref{fig2}(b)]. Note also that the coercivity 'in the dark'
(i.e., recorded for $P = 10$~$\mathrm{\mu W}$) also strongly
depends on temperature, as shown in Fig.~\ref{fig2}(d). To
separate the two contributions, we plot in Fig.~\ref{fig3}(a) the
normalized PCE, i.e., $(H_{cd}-H_{cl}) / H_{cd}$, as a function of
temperature. Clearly, the normalized PCE decreases rapidly with
rising temperature.

Interestingly, this behavior correlates with the thermal
dependence of the resistance $R (T)$ of the sample obtained from a
two terminal measurement between two indium contacts about 3~mm
apart, and presented in Fig.~\ref{fig3}(b). Because of the
relatively low Mn concentration, the sample becomes highly
resistive at low temperatures. We find that the PCE in this sample
only is large when the sample is highly resistive. This may
indicate that the PCE has to do with the inhomogeneous doping
distribution within the layer. Such a conclusion is also
consistent with control experiments we have performed on a higher
doped ($x = 0.05$) perpendicularly magnetized
$\mathrm{Ga_{1-x}Mn_{x}As}$ layer \cite{Noth}. In this high
quality metallic sample, where the resistivity does not change
notably in the temperature range relevant here, we have not
observed any decrease of the coercivity for illumination powers up
to 50~mW.

To further characterize the PCE effect we have performed two-color
experiments. Figure~\ref{fig3}(c) shows field scans of the Kerr
angle probed using a low power HeNe laser beam ($P =
10$~$\mathrm{\mu W}$) under additional illumination from a Xenon
lamp. The white-light radiation from this lamp is dispersed by a
monochromator (set for a maximum intensity centered around a
photon energy $\hbar \omega = 1.96$~eV) and defocused such that
the power density is of the order of $W = 1$~$\mathrm{W \,
cm^{-2}}$. As is clearly seen in the figure, the PCE is detected
even at this low power density: The difference in the coercivity
with and without the illumination from the Xenon lamp is $H_{cd} -
H_{c(\mathrm{Xe})} = 25$~Oe.

This setup allows us to record the spectral dependence of the PCE,
simply by scanning the monochromator. The results of such
experiments is shown in Fig.~\ref{fig3}(d). The solid line in this
figure gives the change in coercivity normalized by the
illumination density $W$ of the Xe lamp, $\frac{
H_{cd}-H_{c(\mathrm{Xe})} }{W}$. (Experimentally, we have observed
that  the PCE is linear with $W$ for small $W$.) For energies
above ca. $1.8$~eV, the normalized PCE increases with photon
energy $\hbar \omega$. We suggest that the spectral response
mainly results from the penetration depth of the photons. For
comparison, we have also plotted the penetration depth $\Lambda
(\hbar \omega) = \alpha ^{-1} (\hbar \omega)$ in nonmagnetic GaAs,
where $\alpha$ is the absorption coefficient, in the same
Fig.~\ref{fig3}(d). At the highest photon energy ($\hbar \omega =
2.41$~eV), we find $\Lambda = 0.1$~$\mathrm{\mu m}$, which is two
and half times shorter than the corresponding $\Lambda$  for
photons from a HeNe laser ($\hbar \omega = 1.96$~eV) and three
times shorter than the layer thickness $d = 0.36$~$\mathrm{\mu
m}$. This means that the PCE occurs within the (Ga,Mn)As layer and
effectively rules out any contribution to the effect arising from
the substrate, the buffer layer, or the GaAs/(Ga,Mn)As interface.

\begin{figure}[tpb]
\includegraphics[width=.37\textwidth]{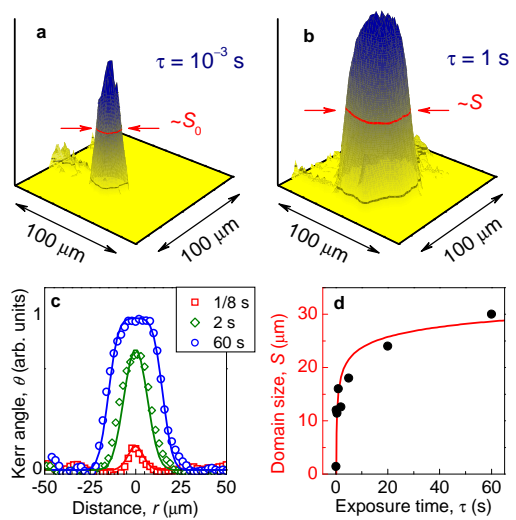}
\caption{The spatially-resolved dynamics of the PCE. (a)
Magnetization profile of a $100 \, \mathrm{\mu m} \times 100 \,
\mathrm{\mu m} $ area obtained after an exposure time $\tau =
10^{-3}$~s (in a magnetic field $H_0 = 480$~Oe) with a focused
beam with $P = 500$~$\mathrm{\mu W}$. (b) The same after exposure
time $\tau = 1$~s. (c) Cross sections of the magnetization
profiles recorded for various exposure times in a magnetic field
$H_0 = 350$~Oe. The axis of the amplitude of the Kerr signal is
calibrated such that 0 and 1 correspond to the Kerr angles
obtained for the layer uniformly magnetized down- and upwards,
respectivly. The solid lines are fits to Eq.~(\ref{Profile}) (d)
Size of the reversed domain $S$ as a function of the exposure time
$\tau$. Solid line is a fit to Eq.~(\ref{Broad}) with $\Delta =
6$~$\mu m$ and $\tau_0 = 0.2$~s. } \label{fig4}
\end{figure}

Figure~\ref{fig4} illustrates the spatially-resolved dynamics of
the PCE, whith the pulse duration controlled by opening time of a
mechanical shutter. The procedure is similar to the MO recording
of Fig.~\ref{fig1}: First, the sample is homogeneously magnetized
in a large magnetic field ($-1$~kOe). Subsequently, the sample is
illuminated for a time period $\tau$ by a focused light beam ($P =
500$~$\mathrm{\mu W}$) in a magnetic field $H_0 = + 480$~Oe. The
profile of the resulting domain of reversed magnetization is then
visualized in zero magnetic field using scanning Kerr microscopy.
For short illumination [$\tau = 10^{-3}$~s, Fig.~\ref{fig4}(a)],
the recorded profile is to a good approximation limited by the
laser spot size, yielding $S_0 \approx 10$~$\mathrm{\mu m}$. For
longer exposure times, the profile widens significantly.
Figure~\ref{fig4}(c) shows linescans of the Kerr signal $\theta
(r)$ through the centre of the illuminated area, characterizing
the gradual increase of the size of the reversed domain with
$\tau$. In order to model these scan profiles we start from our
gaussian laser spot profile $\exp (-r^2 / \Delta^2)$ with $\Delta
= 6$~$\mu m$, which is obtained from the full width at half
maximum $S_0 = 2 \Delta \sqrt{\ln 2}$. The convolution of the
laser spot profile with a circle of diameter $S$ (corresponding to
the size of the reversed domain) yields the Kerr signal $\theta$
as a function of distance $r$. For simplicity we consider the one
dimensional case to be a good approximation and obtain the
analytical solution
\begin{equation}
\theta (r) = \frac{1}{2} \theta_i \left[ \mathrm{erf} \left(
\frac{S/2 + r}{\Delta} \right) + \mathrm{erf} \left( \frac{S/2 -
r}{\Delta} \right) \right] \,. \label{Profile}
\end{equation}
where $\mathrm{erf} ()$ denotes the error function. Fitting to
Eq.~(\ref{Profile}) as shown by solid lines in Fig.~\ref{fig4}(c)
($\theta_i = 1$ for all curves), allows us to extract $S$.

In Fig.~\ref{fig4}(d) we plot $S$ vs. exposure time $\tau$. The
size of the reversed domain increases up to ca. $S =
30$~$\mathrm{\mu m}$ for a $\tau = 60$~s exposure. One could
assume this behavior to be well described in terms of a diffusion
process, i.e. $S/2 \approx \sqrt{D \tau}$. However, substituting
the above numbers would yield a very small diffusion constant of
$D \sim 10$~$\mathrm{\mu m^{2} / s}$. This number is in stark
contrast to the low-temperature thermal conductivity of GaAs which
is five orders of magnitudes larger \cite{Heat_conduct, Heat_GaAs}
and thus effectively rules out any mechanism of the PCE related to
heating effects.



We are now in a position to discuss in some more detail the
mechanism leading to the PCE. With an average Mn concentration of
$x \approx 0.005$, the sample is insulating at low temperatures
[Fig.~\ref{fig3}(b)]. This, combined with the fact that SQUID
shows ferromagnetic behavior at low temperature implies an
inhomogeneous distribution of magnetic ions and holes, creating a
rather irregular potential landscape. Under illumination, the
photogenerated carriers will tend to  smoothen this landscape by
screening effects. This may well result in a lowering of the
domain wall pinning energy - and thus reduce the coercive field.
This mechanism also explains the absence of the effect in the
metallic control sample because in a metallic sample, any local
inhomogeneity will be effectively screened out by mobile holes.
The potential landscape in this sample is thus already smooth
without illumination, and photoexcited carriers have no influence
on domain wall pinning. This argument is also consistent with the
observation that, despite a higher saturation magnetization, the
metallic control sample has a $\sim 3$ times smaller coercivity
than the Mn-doped sample with $x \approx 0.005$.

The proposed model suggests that the local magnetization switching
correlates with reducing the roughness of the disorder potential below some threshold value,
and hence should depend on to the number of photocarriers
generated during illumination. The higher the local intensity
(i.e., closer to the centre of the spot) the shorter is the exposure
time required to induce such a switching. Assuming that the
product of the local intensity and exposure time should be a
constant, one obtains for the lateral broadening of the
remagnetization area
\begin{equation}
S = 2 \Delta \sqrt{\ln \frac{\tau}{\tau_0}} \,. \label{Broad}
\end{equation}
This reasonably reproduces the experimental behavior of
Fig.~\ref{fig4}(d) with $\tau_0 = 0.2$~s (solid line in the
figure). The physical meaning of $\tau_0$ is the time required to
optically create the smallest magnetic domain. Obviously, $\tau_0$
should depend on the illumination power $P$ and switching magnetic
field $H_0$.


Summarizing, we have demonstrated local manipulation of the
coercive field in a nearly insulating (Ga,Mn)As sample by low
intensity illumination and performed a complete recording cycle
(i.e., initialization, writing, deleting, and rewriting) on a
lateral scale of a few microns. We suggest a disorder related
mechanism as the most likely cause for this PCE. This proof of
concept demonstration for low power MO recording is based on
focussing the action of the light on modifying the pinning centers
which control domain wall movement, rather than on the tradition
approach of modifying the magnetization strength.


\begin{acknowledgments}

We thank M.~Sawicki from IFPAN, Poland for useful discussions and
R.~P.~Campion, A.~W.~Rushforth and B.~L.~Gallagher at Nottingham
University, U.K. for providing the high quality metallic sample
used in the control experiment. This work was supported by
DFG-RFBR and RSSF.

\end{acknowledgments}

\end{document}